\documentclass[a4paper]{jpconf}
\usepackage{lineno,graphicx}
\DeclareUnicodeCharacter{2212}{\ensuremath{-}}
\begin{document}
\title{Hadronic resonance production with ALICE at the LHC}

\author{Sergey Kiselev$^*$ for the ALICE Collaboration}

\address{$^*$NRC "Kurchatov institute", Moscow, Russia}

\ead{Sergey.Kiselev@cern.ch}

\begin{abstract}
Recent results on short-lived hadronic resonances obtained by the ALICE experiment at LHC energies are presented. 
Results include  system-size  and  collision-energy  
evolution of transverse momentum spectra, yields and ratios of resonance yields 
to those of longer lived particles, and nuclear modification factors. 
The results are compared with model predictions and measurements at lower energies.
\end{abstract}

Hadronic resonance production plays an important role both in elementary and in nucleus-nucleus collisions.
In heavy-ion collisions, since the lifetimes of short-lived resonances are comparable with 
the lifetime of the late hadronic phase, regeneration and rescattering effects become 
important and ratios of yields of resonances relative to those of longer lived particles can be used to estimate 
the time interval between the chemical and kinetic freeze-out. 
The measurements in pp and p--Pb collisions constitute a reference for nuclear collisions 
and provide information for tuning event generators inspired by Quantum Chromodynamics. 

Results on short-lived 
mesonic $\rho(770)^{0}$, $\mathrm{K}^{*}(892)^{0}$, $\mathrm{K}^{*}(892)^{\pm}$, $f_{0}(980)$, $\phi(1020)$ 
as well as baryonic $\Sigma(1385)^{\pm}$, $\Lambda(1520)$ and $\Xi(1530)^{0}$ resonances 
(hereafter $\rho^{0}$, $\mathrm{K}^{*0}$, $\mathrm{K}^{*\pm}$, $f_{0}$, $\phi$, $\Sigma^{*\pm}$, $\Lambda^{*}$, $\Xi^{*0}$) 
have been obtained using data reconstructed with the ALICE detector.
The  resonances  are  reconstructed  in  their  hadronic  decay  channels  and have very different lifetimes
as shown in Table~\ref{tab:Res}. 
\begin{table}[ht]
\caption{Reconstructed decay mode and lifetime values~\cite{PDG} for hadronic resonances.}
\begin{center}
\begin{tabular}{ l l l l l l l l l}
\br
                    &$\rho^{0}$&$\mathrm{K}^{*0}$  &$\mathrm{K}^{*\pm}$  & $f_{0}$ &  $\Sigma^{*\pm}$ &    $\Lambda^{*}$    &     $\Xi^{*0}$   &  $\phi$    \\
\mr
decay channel&$\pi\pi $&$\mathrm{K}\pi $&$\mathrm{K^{0}_{S}}\pi $&$\pi\pi $&$\Lambda\pi $&$p\mathrm{K} $&$\Xi\pi $&$\mathrm{K}\mathrm{K} $\\
lifetime (fm/\it{c})& 1.3 & 4.2 & 4.2 & $\sim$ 5& 5-5.5 & 12.6 & 21.7& 46.2 \\ 
ALICE papers &\cite{ALICErho0}&\cite{ALICEpp7}-\cite{ALICEppXe} &\cite{ALICEppKstar-pm}-\cite{ALICEPb5Kstar-pm} &\cite{ALICEppf0} &\cite{ALICEppSigmaStar}-\cite{ALICEPbPbSigmaStar} &\cite{ALICELambdaStar}-\cite{ALICELambdaStar2} &\cite{ALICEppSigmaStar}-\cite{ALICEpp13SstarXstar} &\cite{ALICEpp7}-\cite{ALICEpPb502} \\
\br
\end{tabular}
\end{center}
 \label{tab:Res}
\end{table}
%
This contribution reports recent results obtained 
for $\mathrm{K}^{*0}$ and $\phi$ in p--Pb at $\sqrt{s_{\mathrm {NN}}}$ =5.02 \cite{ALICEpPb502} and 8.16 TeV \cite{ALICEpPb816}, pp and Pb--Pb at 5.02 TeV \cite{ALICEppPbPb502},
for $\mathrm{K}^{*\pm}$ in pp at 5.02, 8, 13 TeV \cite{ALICEppKstar-pm} and Pb--Pb at 5.02 TeV \cite{ALICEPb5Kstar-pm},
for $f_{0}$ in pp \cite{ALICEppf0} and p--Pb at 5.02 TeV,
for $\Sigma^{*\pm}$ in Pb--Pb at 5.02 TeV \cite{ALICEPbPbSigmaStar},
for $\Lambda^{*}$ in pp at 5.02, 13 TeV and Pb--Pb at 5.02 TeV.

\begin{figure}[hbtp]
\begin{center}
\includegraphics[scale=0.7]{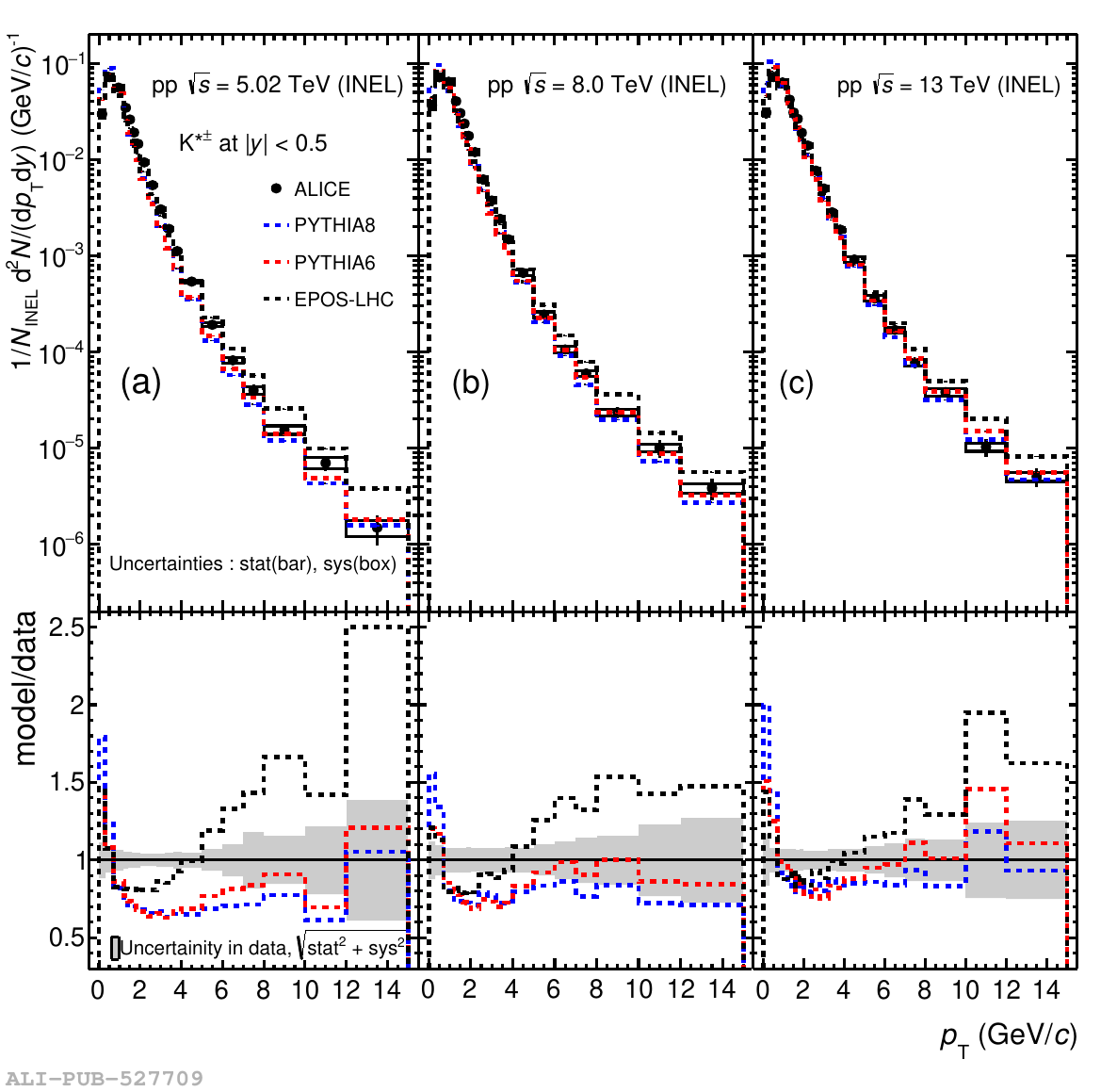}
\end{center}
\caption{(color online) 
The $\mathrm{K}^{*\pm}$ $p_{\rm T}$ spectra (black dots) measured in inelastic pp collisions at (a) $\sqrt{s}$ = 5.02 TeV, (b) 8 TeV, and (c) 13 TeV 
are compared to the distributions predicted by PYTHIA6~\cite{Perugia}, PYTHIA8~\cite{Monash}, and EPOS-LHC~\cite{EPOSLHC}. 
}
  \label{fig:spectra}
\end{figure}

Figure~\ref{fig:spectra} shows the transverse momentum spectra for $\mathrm{K}^{*\pm}$ in inelastic pp collisions at $\sqrt{s}$ = 5.02, 8, and 13.
%
Data are compared with PYTHIA6, PYTHIA8 and EPOS-LHC predictions. 
The ability of the models to both qualitatively and quantitatively describe the data improves with the collision energy. 
The best agreement is obtained with PYTHIA for 13 TeV.

\begin{figure}[hbtp]
\begin{center}
\includegraphics[scale=0.39]{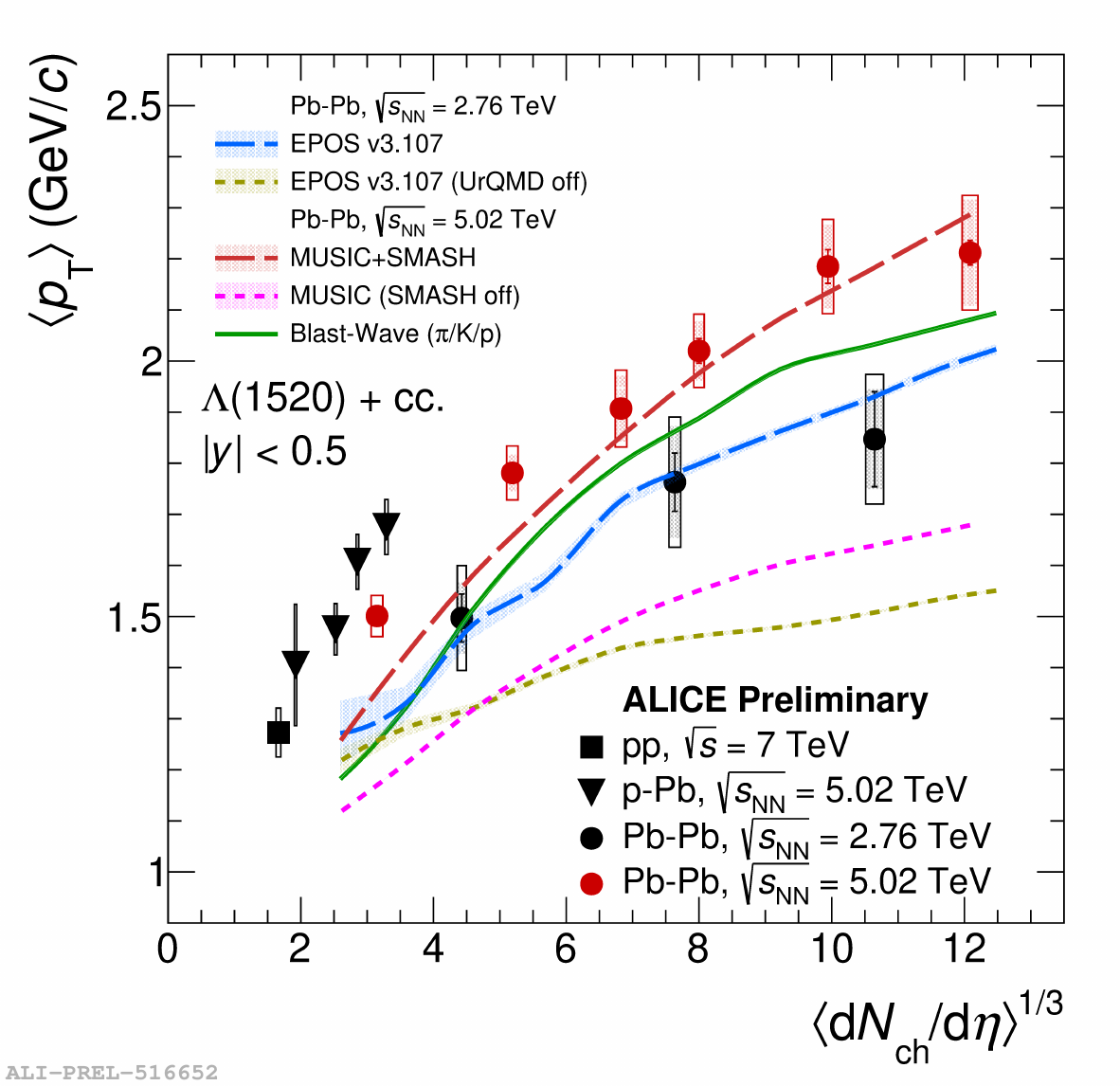}
\includegraphics[scale=0.39]{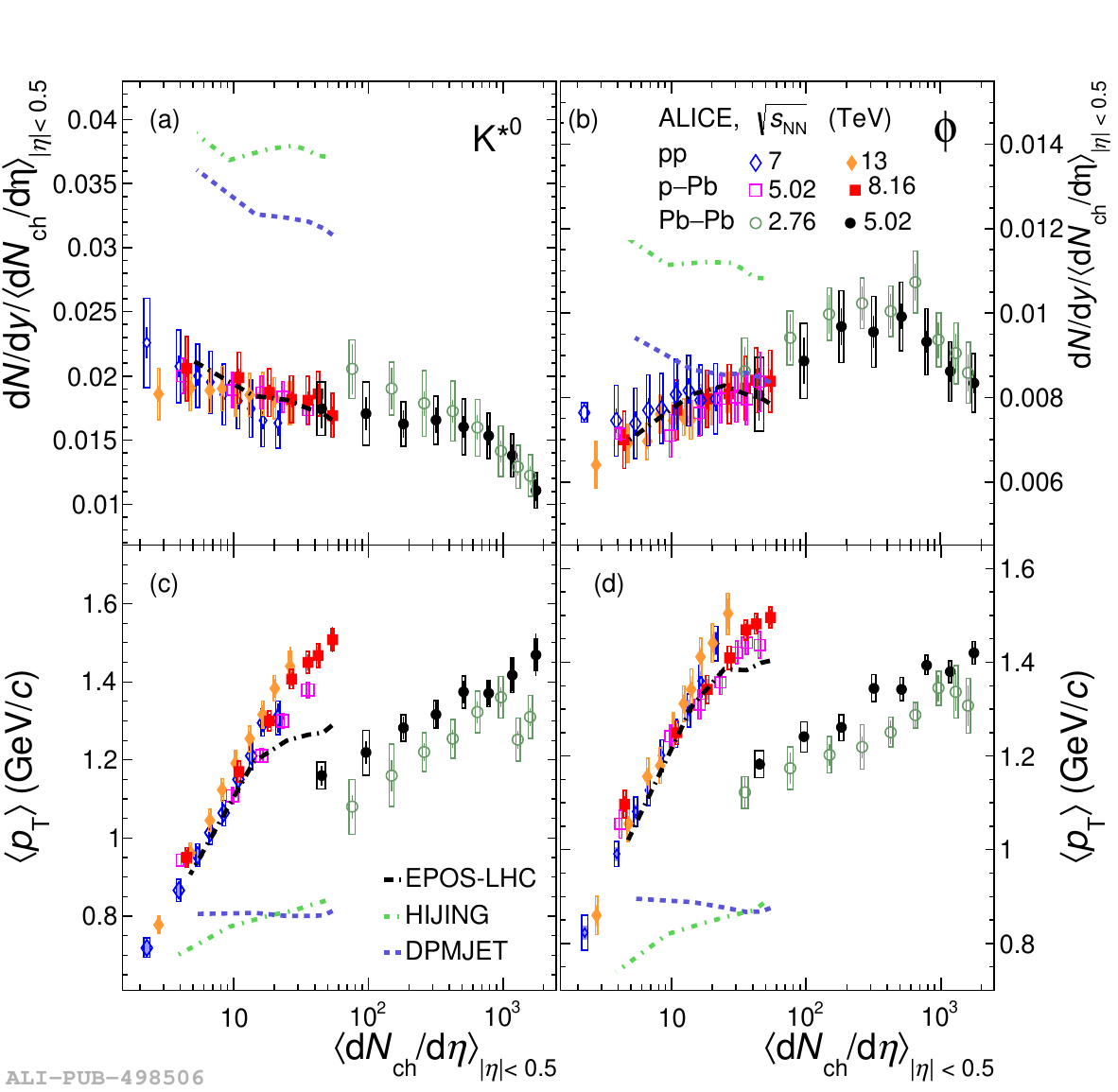}
\end{center}
\caption{(color online) 
Left: The mean transverse momentum of $\Lambda^{*}$ as a function of the charged-particle multiplicity density in Pb--Pb collisions at $\sqrt{s_{\mathrm {NN}}}$ = 5.02 TeV with model predictions from Blast-Wave \cite{BlastWave-piKp} and MUSIC \cite{MUSIC} with and without the hadronic phase (SMASH off). Results are also shown for pp at $\sqrt{s}$ = 7 TeV \cite{ALICELambdaStar2}, p--Pb at $\sqrt{s_{\mathrm {NN}}}$ = 5.02 TeV \cite{ALICELambdaStar2} and Pb--Pb at $\sqrt{s_{\mathrm {NN}}}$ = 2.76 TeV with EPOS3 predictions \cite{ALICELambdaStar}.
Right: The multiplicity-scaled $p_{\rm T}$-integrated yield (a, d) and mean transverse momentum (c, d) of $\mathrm{K}^{*0}$(a, c) and $\phi$(b, d) 
as functions of the charged-particle multiplicity density in p--Pb collisions at $\sqrt{s_{\mathrm {NN}}}$ = 8.16 TeV  with model predictions from EPOS-LHC \cite{EPOSLHC}, 
DPMJET \cite{DPMJET} and HIJING \cite{HIJING}. The measurements in pp at $\sqrt{s}$ = 7 \cite{ALICEpp7}, 13 \cite{ALICEpp13} TeV, p--Pb 
at $\sqrt{s_{\mathrm {NN}}} = 5.02$ TeV \cite{ALICEpPb} and Pb--Pb at $\sqrt{s_{\mathrm {NN}}} = 2.76$ \cite{ALICEPbPb}, 5.02 \cite{ALICEPbPb502} TeV 
are also shown. 
}
  \label{fig:mpt}
\end{figure}

The average transverse momentum of $\Lambda^{*}$ as a function of the charged-particle multiplicity density measured in pp, p--Pb and Pb--Pb collisions 
is shown in Fig.~\ref{fig:mpt} (left).
%
New data from Pb--Pb collisions at $\sqrt{s_{\mathrm {NN}}}$ = 5.02 TeV show larger values than at $\sqrt{s_{\mathrm {NN}}}$ = 2.76 TeV and predictions from the Blast-Wave model.
Predictions from the MUSIC+SMASH model, which includes modeling of the hadronic phase, are consistent with the data in central collisions but underestimate them in peripheral ones.
Models without hadronic afterburner underestimate the measurements at both energies.
Figure~\ref{fig:mpt} (right, bottom) shows the mean transverse momentum of $\mathrm{K}^{*0}$(c) and $\phi$(d) as a function of the charged-particle 
multiplicity density. New data values for p--Pb at $\sqrt{s_{\mathrm {NN}}}$ = 8.16 TeV are close to the values for other light systems.
Among the models, EPOS-LHC gives the best agreement with the data. 
In pp and p--Pb collisions, the $\langle p_\mathrm{T}\rangle$ rises faster with multiplicity than in Pb--Pb collisions.
An analogous behavior has been observed in~\cite{ALICE_mpt} for charged particles and can be understood as the effect of color reconnection 
between strings produced in multi-parton interactions. 

Normalized $p_{\rm T}$-integrated yields of $\mathrm{K}^{*0}$ and $\phi$ as a function of the charged-particle multiplicity density are presented 
in Fig.~\ref{fig:mpt} (right, top). New data for p--Pb collisions at $\sqrt{s_{\mathrm {NN}}}$ = 8.16 TeV follow the general trend: 
yields are independent of collision system and energy and appear to be driven by the event multiplicity.
EPOS-LHC predictions agree with the data.

\begin{figure}[hbtp]
\begin{center}
\includegraphics[scale=0.42]{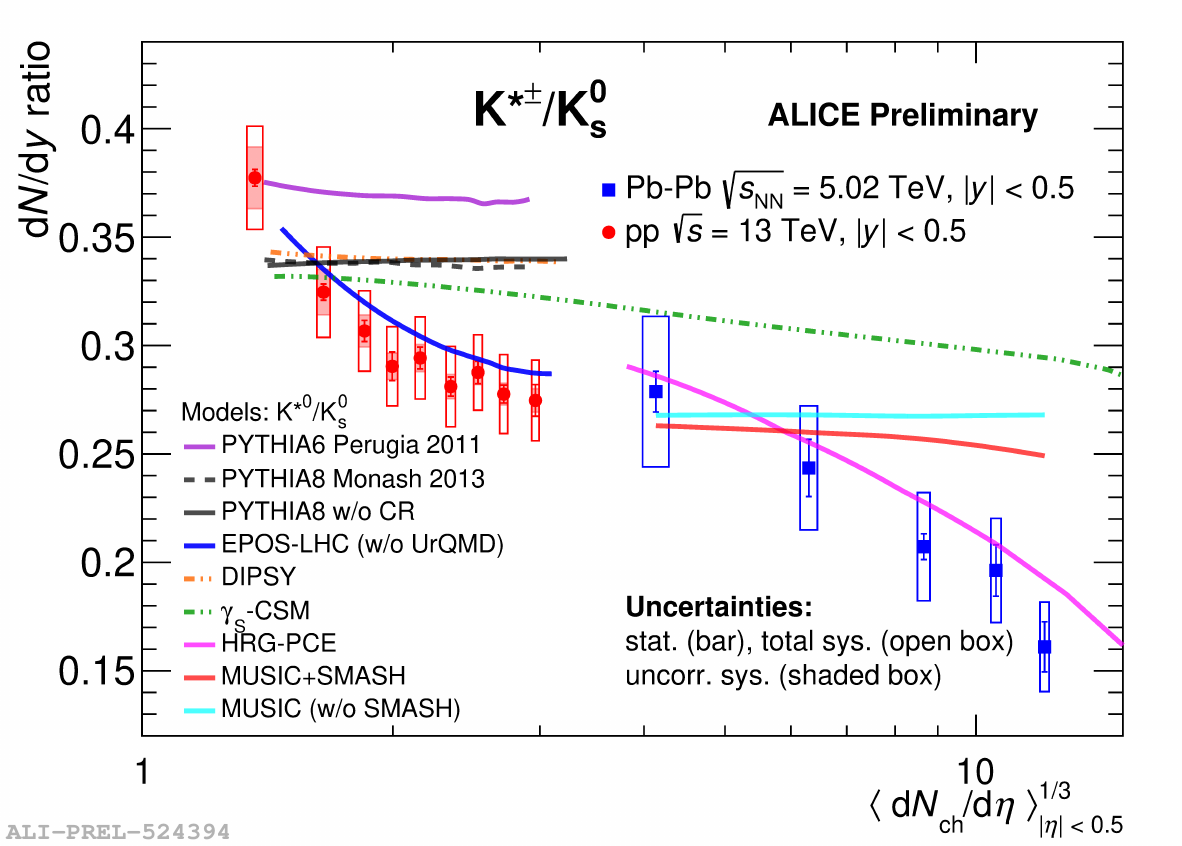}
\includegraphics[scale=0.34]{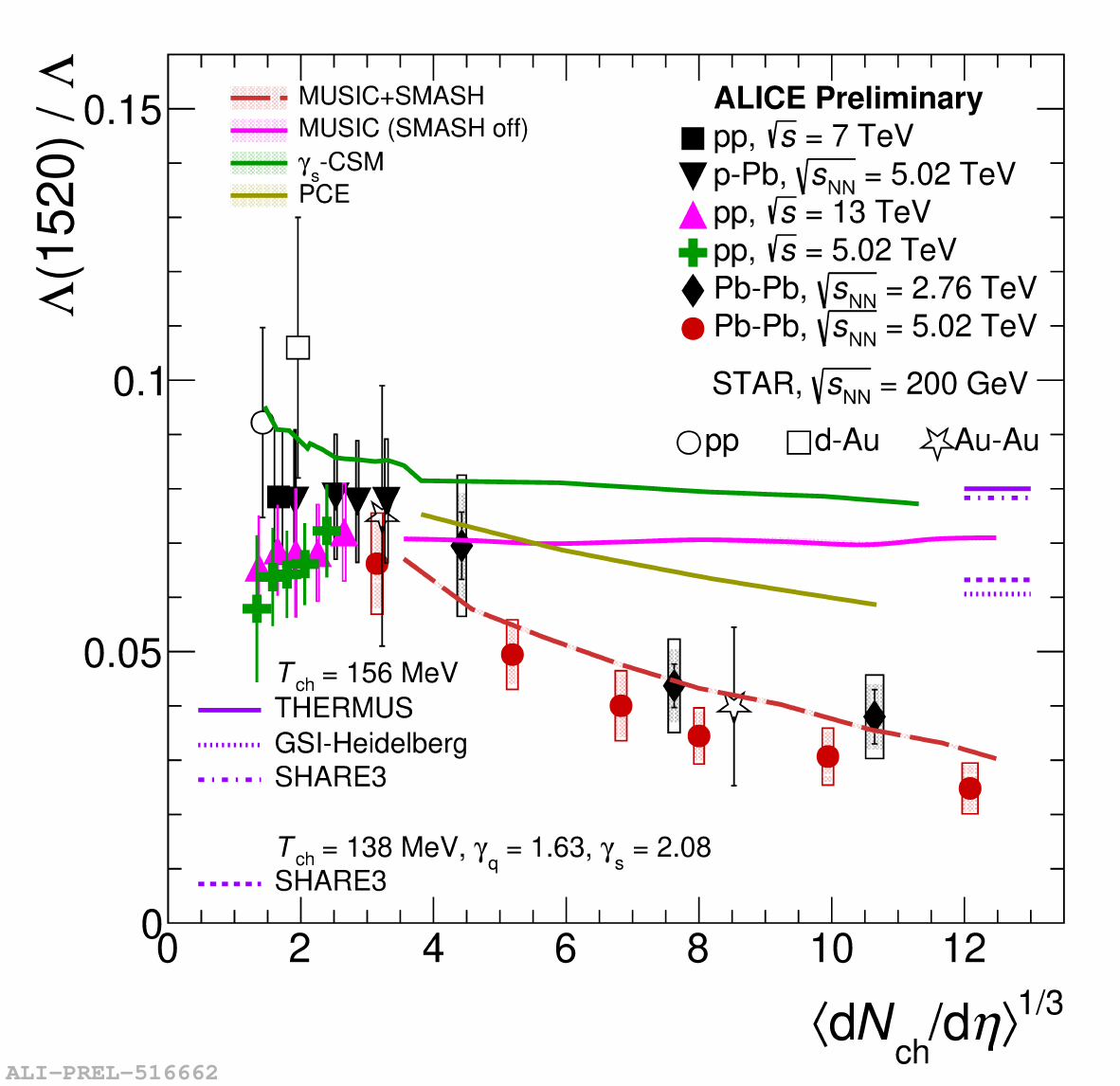}
\end{center}
\caption{(color online) 
Particle yield ratios $\mathrm{K}^{*\pm}/\mathrm{K^{0}_{S}}$ (left) and $\Lambda^{*}/\Lambda$ (right) as a function of the charged-particle 
multiplicity density in pp, p--Pb and Pb--Pb collisions  with model predictions from PYTHIA6 \cite{Perugia}, PYTHIA8 \cite{Monash}, 
EPOS-LHC \cite{EPOSLHC}, DIPSY \cite{DIPSY}, $\gamma_{S}$-CSM \cite{GsCSM}, HRG-PCE \cite{HRG-PCE}, MUSIC \cite{MUSIC}, THERMUS \cite{THERMUS}, 
GSI-Heidelberg \cite{GSIHeidelberg} and SHARE3 \cite{SHARE}. STAR data \cite{STAR} are also shown for $\Lambda^{*}/\Lambda$ ratios.
}
  \label{fig:Mratios}
\end{figure}

Figure~\ref{fig:Mratios} (left) shows the particle yield ratios $\mathrm{K}^{*\pm}/\mathrm{K^{0}_{S}}$ as a function of 
the charged-particle multiplicity density.
The ratio is significantly suppressed, by about 55\%, going from peripheral to central Pb--Pb collisions.
This suppression is consistent with rescattering of $\mathrm{K}^{*\pm}$ decay products in the hadronic phase of central collisions as the dominant effect.
Models with rescattering effect (MUSIC+SMASH and HRG-PCE) qualitatively describe the data.
There is a hint of decrease of the ratio with increasing multiplicity in pp collisions.
Among the models, EPOS-LHC gives the best agreement with the data.
The $\mathrm{K}^{*\pm}$ measurement is consistent with previous results for $\mathrm{K}^{*0}$ \cite{ALICEpp13}.
It is worth noting that the systematic uncertainties of charged $\mathrm{K}^{*}$ are smaller than those of neutral $\mathrm{K}^{*}$. 

The particle yield ratios $\Lambda^{*}/\Lambda$ are illustrated in Fig.~\ref{fig:Mratios} (right).
The new measurements in Pb--Pb collisions at $\sqrt{s_{\mathrm {NN}}}$ = 5.02 TeV follow the suppression trend observed for
Pb--Pb at $\sqrt{s_{\mathrm {NN}}}$ = 2.76 TeV \cite{ALICELambdaStar}. 
The suppression, of about 70\%, is larger than for 
$\mathrm{K}^{*\pm}/\mathrm{K^{0}_{S}}$ although $\tau(\Lambda^{*}) \approx 3 \tau(\mathrm{K}^{*\pm})$.
The difference might also be explained by a difference in the interaction cross sections
of the resonance decay products with pions, the most abundant species in the hadron gas.
The model MUSIC with SMASH afterburner reproduces the suppression trend. 
All thermal models overestimate the ratio in central Pb--Pb collisions.
A multiplicity-dependent suppression is not observed in the new data for pp collisions at $\sqrt{s}$ = 5.02 and 13 TeV.
ALICE results confirm the trend seen by STAR at 200 GeV.

\begin{figure}[hbtp]
\begin{center}
\includegraphics[scale=0.36]{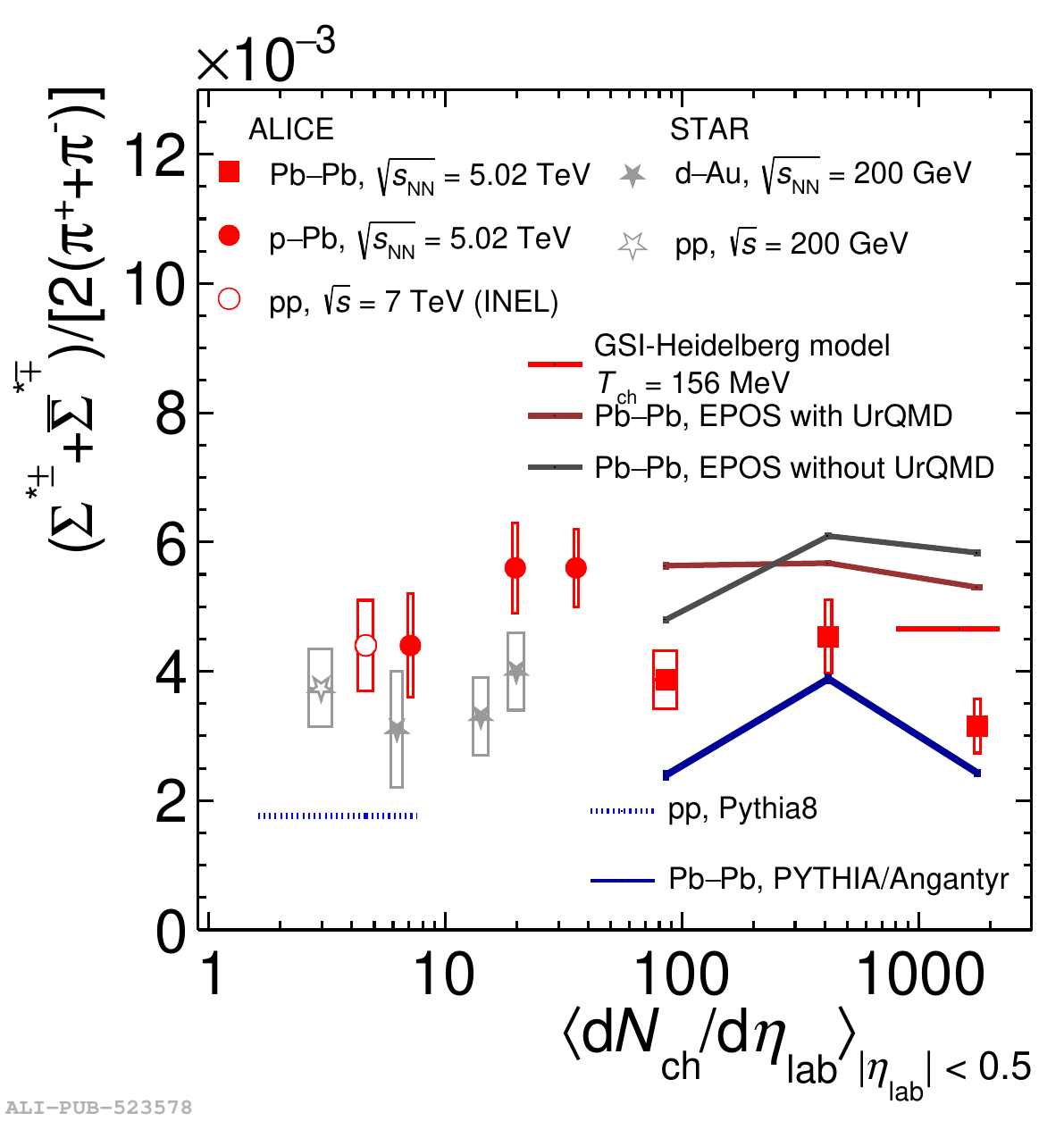}
\includegraphics[scale=0.41]{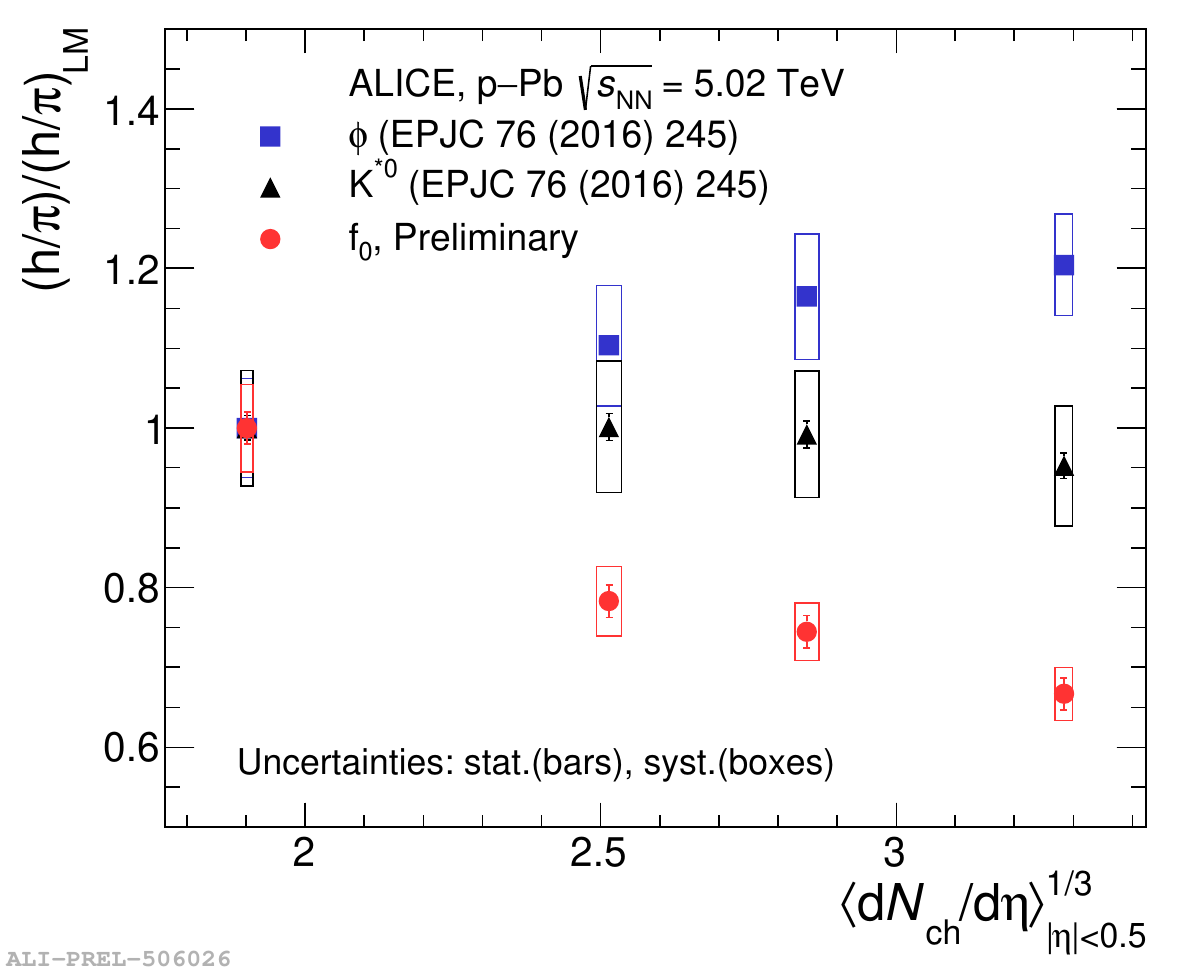}
\end{center}
\caption{(color online)
Left: Particle yield ratios $\Sigma^{*\pm}/\pi$ as a function of the charged-particle multiplicity density in pp, p--Pb and Pb--Pb collisions  
with model predictions from PYTHIA/Angantyr \cite{PYTHIA8Angantyr}, EPOS \cite{KnospeEPOS} and  GSI-Heidelberg \cite{GSIHeidelberg}. 
STAR data \cite{STAR} are also shown. 
Right: Double ratio of particle yields to pion yields as a function of multiplicity in p--Pb collisions at $\sqrt{s_\mathrm{NN}}$  = 5.02 TeV.
}
  \label{fig:f0}
\end{figure}

Figure~\ref{fig:f0} (left) shows the particle yield ratios $\Sigma^{*\pm}/\pi$ as a function of the charged-particle multiplicity density in pp, p--Pb and Pb--Pb collisions.
New results for Pb--Pb at \mbox{$\sqrt{s_{\mathrm {NN}}}$ = 5.02 TeV} do not exhibit particular trend 
with multiplicity \cite{ALICEPbPbSigmaStar}. There is a hint of some suppression at the highest multiplicity but
future higher precision measurements are needed.
EPOS with UrQMD and PYTHIA/Angantyr models reproduce qualitatively multiplicity dependence.
The GSI-Heidelberg thermal model overestimates the ratio in central Pb--Pb collisions.
In pp and p--A collisions, ALICE results are close to STAR data. 

To investigate the structure of $f_{0}$, the new ALICE measurements of the $f_{0}$ yields
are compared to those of other hadrons and resonances with similar mass.
Figure~\ref{fig:f0} (right) shows the double ratio of particle yields to pion yields as a function of multiplicity 
in p--Pb collisions at $\sqrt{s_\mathrm{NN}}$  = 5.02 TeV.
Strangeness enhancement with multiplicity could explain the increase with multiplicity of the $\phi/\pi$ ratios.
The $\mathrm{K}^{*0}/\pi$ ratios demonstrate the competition between strangeness
enhancement and rescattering effect. 
The $f_{0}/\pi$ suppression shows that rescattering is the effect that dominantly affects 
the yield at low $p_{\rm T}$. 
Notably, the $\gamma_{S}-$CSM model prediction for the $f_{0}/\pi$ ratio assuming zero net strangeness of $f_{0}$ 
is consistent with the data within 1.9$\sigma$ ~\cite{ALICEppf0}.

\begin{figure}[hbtp]
\begin{center}
\includegraphics[scale=0.7]{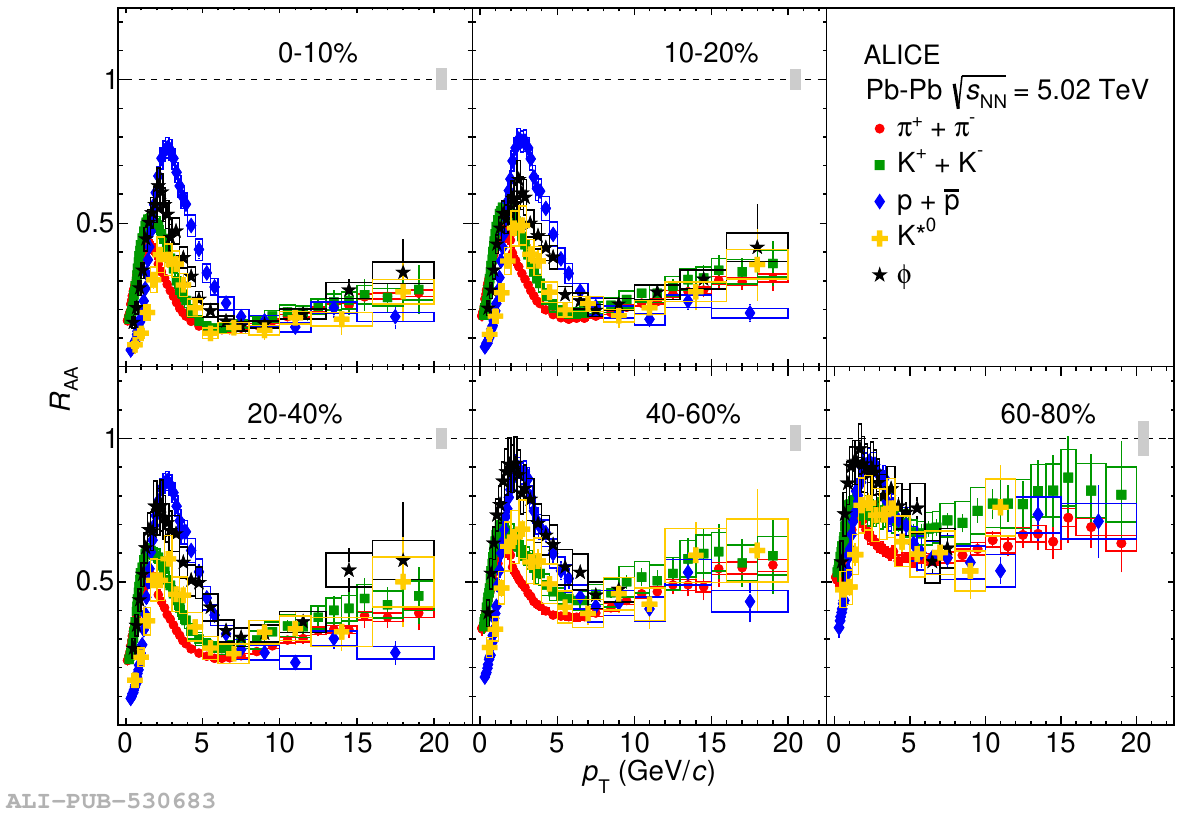}
\end{center}
\caption{(color online) The nuclear modification factor $R_{\rm AA}$ for $\mathrm{K}^{*0}$ and $\phi$ as a function of transverse momentum in Pb--Pb collisions 
for different centrality classes at $\sqrt{s_\mathrm{NN}}$  = 5.02 TeV. The results are compared with $R_{\rm AA}$ for $\pi$, K and p \cite{RAApiKp}. 
}
  \label{fig:RAA}
\end{figure}

Figure~\ref{fig:RAA} presents the nuclear modification factor $R_{\rm AA}$ of $\mathrm{K}^{*0}$ and $\phi$ for Pb--Pb collisions at $\sqrt{s_\mathrm{NN}}$  = 5.02 TeV. 
At $p_{\rm T}$ $>$ 8 GeV/c $R_{\rm AA}$ values for $\mathrm{K}^{*0}$, $\phi$ and light-flavored hadrons are similar within uncertainties.
At low $p_{\rm T}$ ($<$ 5 GeV/c), $R_{\rm AA}$ values for $\mathrm{K}^{*0}$ are lower with respect to those obtained for $\phi$, 
chiefly because of rescattering effects.
As demonstrated in ~\cite{ALICEppPbPb502}, new results at $\sqrt{s_{\mathrm {NN}}}$ = 5.02 TeV are comparable to the corresponding measurements 
at \mbox{$\sqrt{s_{\mathrm {NN}}}$ = 2.76 TeV} ~\cite{ALICEPbPb-highPT}.

\begin{figure}[hbtp]
\begin{center}
\includegraphics[scale=0.8]{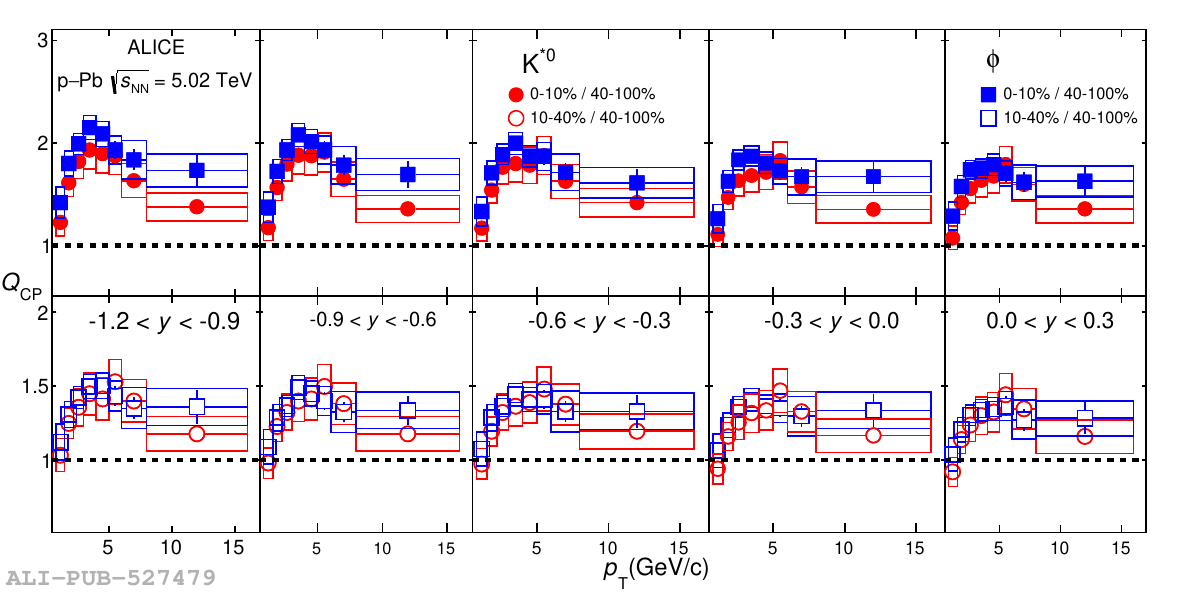}
\end{center}
\caption{(color online) The $Q_{\rm CP}$ of $\mathrm{K}^{*0}$ (red circles) and $\phi$ (blue squares) mesons as a function of $p_{\rm T}$ 
for 0-10\%/40-100\% (top panels) and 10-40\%/40-100\% (bottom panels) in various rapidity intervals within the range
−1.2 $<$ y $<$ 0.3 in p--Pb collisions at $\sqrt{s_\mathrm{NN}}$ = 5.02 TeV. 
}
  \label{fig:QCP}
\end{figure}

Multiplicity and rapidity dependence of the nuclear modification factor $Q_{\rm CP}$ for $\mathrm{K}^{*0}$ and $\phi$ mesons in p--Pb collisions 
at $\sqrt{s_\mathrm{NN}}$  = 5.02 TeV is presented in Fig.~\ref{fig:QCP}.
$Q_{\rm CP}$ values as a function of $p_{\rm T}$ show a bump, with a maximum around 3 GeV/c, suggestive of the Cronin effect. This Cronin-like enhancement 
is more pronounced for large negative rapidities (in the Pb-going direction) and for more central (higher multiplicity) collisions.
 
In summary, recent results on short-lived hadronic resonances obtained by the ALICE experiment in pp, p--Pb and Pb--Pb collisions at the LHC energies have been presented.
In pp and p--Pb collisions the $\langle p_\mathrm{T}\rangle$ values for the $\mathrm{K}^{*0}$ and $\phi$ resonances 
rise faster with multiplicity than in Pb--Pb collisions.
One possible explanation could be the effect of color reconnection between strings produced in multi-parton interactions.
Yields of $\mathrm{K}^{*0}$ and $\phi$ are independent of the collision system and energy and appear to be driven by the event multiplicity.
The  $\mathrm{K}^{*\pm}/\mathrm{K^{0}_{S}}$,  and $\Lambda^{*}/\Lambda$ ratios exhibit a significant suppression 
going from peripheral to central \mbox{Pb--Pb} collisions, consistent with rescattering of the decay products of the short-lived resonances in the hadronic phase.
There is a hint of suppression for the $\Sigma^{*\pm}/\pi$ ratio but future higher precision measurements are needed.
The suppression is qualitatively described by models with rescattering.
For light systems there are hints of suppression for $\mathrm{K}^{*\pm}/\mathrm{K^{0}_{S}}$ and $f_{0}/\pi$ ratios
and no suppression for $\Lambda^{*}/\Lambda$.  
%
In Pb--Pb collisions $R_{\rm AA}$ for $\mathrm{K}^{*0}$, $\phi$ and light-flavored hadrons are similar within uncertainties at $p_{\rm T} >$ 8 GeV/c.
In p--Pb collisions $Q_{\rm CP}$ shows Cronin-like enhancement which is more pronounced for large negative rapidities (in the Pb-going direction) 
and for more central (higher multiplicity) collisions.

The work was carried out within the state assignment of NRC "Kurchatov institute".

\end{document}